\documentclass[preprint,12pt]{elsarticle}
\usepackage[utf8]{inputenc}
\usepackage[T1]{fontenc}
\usepackage[margin=1in]{geometry}
\usepackage{lipsum}
\usepackage{xcolor}
\usepackage{listings}
\usepackage{amssymb}
\usepackage{graphicx}
\usepackage{lineno}
\usepackage{subfigure}
\usepackage{siunitx}
\usepackage[version=4]{mhchem}
\usepackage[hidelinks]{hyperref}

\newcommand{\arxiv}{ar\protect\raisebox{0.4ex}{$\chi$}iv}

\journal{\arxiv}

\usepackage{wasysym}
\usepackage{pict2e}
\DeclareRobustCommand{\slashcirc}{{\mathpalette\doslashcirc\relax}}
\makeatletter
\newcommand\doslashcirc[2]{%
  \sbox\z@{$#1\m@th\circ$}%
  \setlength\unitlength{\wd\z@}
  \begin{picture}(1,1)
  \roundcap
  \put(0,0){\box\z@}
  \put(0,0){\line(1,1){1}}
  \end{picture}%
}
\makeatother

\begin{document}
\begin{frontmatter}
\title{Neuromorphic Liquid Marbles With Aqueous Carbon Nanotube Cores}

\author{Richard Mayne$^{a,b}$}
\ead{Richard.Mayne@uwe.ac.uk}
\author{Thomas C.\ Draper$^b$}
\author{Neil Phillips$^b$}
\author{James G.\ H.\ Whiting$^{c,b,d}$}
\author{Roshan Weerasekera$^{c,b}$}
\author{Claire Fullarton$^b$}
\author{Ben P.\ J.\ de Lacy Costello$^{a,b}$}
\author{Andrew Adamatzky$^b$}

\address[das]{Department of Applied Sciences, Faculty of Health and Applied Sciences, \mbox{University of the West of England, Frenchay Campus, BS16 1QY, UK}}
\address[ucg]{Unconventional Computing Group, Faculty of the Environment and Technology, \mbox{University of the West of England, Frenchay Campus, BS16 1QY, UK}}
\address[edm]{Department of Engineering Design and Mathematics, Faculty of the Environment and Technology, \mbox{University of the West of England, Frenchay Campus, BS16 1QY, UK}}
\address[hth]{Health Technology Hub, University of the West of England, Frenchay Campus, BS16 1QY, UK}

\begin{abstract}
Neuromorphic computing devices attempt to emulate features of biological nervous systems through mimicking the properties of synapses, towards implementing the emergent properties of their counterparts, such as learning. Inspired by recent advances in the utilisation of liquid marbles (microlitre quantities of fluid coated in hydrophobic powder) for the creation of unconventional computing devices, we describe the development of liquid marbles with neuromorphic properties through the use of copper coatings and \SI{1.0}{\milli\gram\per\milli\litre} carbon nanotube-containing fluid cores. Experimentation was performed through sandwiching the marbles between two cup-style electrodes and stimulating them with repeated DC pulses at \SI{3.0}{\volt}. Our results demonstrate that `entrainment' of a carbon nanotube filled-copper liquid marble via periodic pulses can cause their electrical resistance to rapidly switch between high to low resistance profiles, upon inverting the polarity of stimulation: the reduction in resistance between high and low profiles was approximately 88\% after two rounds of entrainment. This effect was found to be reversible through reversion to the original stimulus polarity and was strengthened by repeated experimentation, as evidenced by a mean reduction in time to switching onset of 43\%. These effects were not replicated in nanotube solutions not bound inside liquid marbles. Our electrical characterisation also reveals that nanotube-filled liquid marbles exhibit pinched loop hysteresis IV profiles consistent with the description of memristors. We conclude by discussing the applications of this technology to the development of unconventional computing devices and the study of emergent characteristics in biological neural tissue. 

\end{abstract}

\end{frontmatter}

\section{Introduction}
Computation is a ubiquitous property of natural matter that, through a universal and objective language, will unite the sciences. More generally, physical systems may be applied to mathematical problems to create machines and computers. Complex systems may be correspondingly abstracted in algorithmic terms in order to describe phenomena that have traditionally evaded the grasp of understanding, such as complexity arising from biological sensorial-actuation networks, through which phenomena such as `intelligence' are hypothesised to emerge \cite{Priel2010,Mayne2014cib,Adamatzky2015actinautomata,Mayne2014,Woolf2010,Hameroff1987,Lahoz-Beltra1993}, even in organisms that do not possess nervous systems \cite{Whiting2016}. This application of computing concepts and development of experimental devices therein encompasses the field of `unconventional computing'.

A neuromorphic characteristic of an engineered system is so named if it mimics the structure or functionality of a component\slash multiple components of the Metazoan nervous system. Typically, this will involve attempts to replicate the phenomenon of synaptic plasticity: self-modulation of the excitability of neuron-neuron junctions (synapses), towards replicating state retention (`learning') via a process of entrainment with graduated input (`neuromodulation'). Neuromorphic devices are worthy of research attention as an unconventional computing paradigm, due to certain features of their biological counterparts --- such as massive parallelism, emergence, and low energy consumption --- being highly desirable to emulate.

This paper aims to create Neuromorphic computing devices from Liquid marbles (LMs). LMs are spherical microlitre quantities of fluid with a superhydrophobic particulate coating, which can range in size between tens and thousands of micrometres in diameter \cite{aussillous2001liquid,McHale2011,McHale2015,bormashenko2016liquid}. These systems exhibit novel characteristics such as low coefficients of friction \cite{aussillous2006properties,bormashenko2009mechanism,bormashenko2010nature}, which have been exploited by nature~\cite{Pike2002,Kasahara2019}. It has been demonstrated that liquid marbles have myriad uses ranging from micro-bioreactors~\cite{bormashenko2011liquid,bormashenko2016liquid,ooi2015manipulation,oliveira2017potential} to gas biosensors~\cite{Tian2010,Tian2010a} to unconventional computing media~\cite{draper2017liquid,draper2018liquid}. Our laboratory has developed LM devices that are capable of implementing computation through a variety of non-standard logics \cite{draper2017liquid,draper2018liquid,draper2018ucnc} where the LMs are considered as data or otherwise, to contain data (i.e.\ chemical reactants), which may interact with other LMs via collisions that will result in data translation or transfer via ricochets or coalescence. Towards these goals, we (and others) have also examined LM dynamics to enhance their usefulness for these purposes, e.g.\ evaporation~\cite{Fullarton2018,Dandan2009} and ballistic interactions~\cite{Draper2019collide,Planchette2013}. The current work is therefore presented as a route towards developing microlitre-quantity three-dimensional ballistic-chemical reactors, that exhibit neuromorphic properties and may hence be used as unconventional computing media. 

To engineer a neuromorphic effect in our LMs, the core chosen was an aqueous dispersion of carbon nanotubes (CNTs). In 2001 Cui {\it et al.}~\cite{cui2002carbon} experimentally demonstrated that single walled-CNTs can be switched between two conductance states (high-conductance and low-conductance), which differ by more than two orders of magnitude, with a threshold voltage shift of \SI{1.25}{\volt}. Theoretical analysis has shown that CNTs can act as Schottky barrier transistors~\cite{heinze2002carbon} and several patent applications for CNT switching devices have been filed~\cite{bertin2006nanotube,choi2006memory}. Regarding progress towards implementations of CNT computing systems, field effect transistors have been described~\cite{avouris2002molecular}, and experimental laboratory evidence suggests that the solid-state switching signatures of CNTs might be due to their relative  mechanical  movements~\cite{diehl2003single}. Carbon nanotube artificial synapses have previously been prototyped separately by K.\ Kim \textit{et al.}~\cite{kim2013carbon} and S.\ Kim \textit{et al.}~\cite{kim2017pattern}: the synapse operates via dynamic interactions between CNTs and  hydrogen  ions  in  an  electrochemical  cell  integrated  in the  synapse. Our aims therefore were to capitalise on these properties of CNTs within LMs.

The following paper is structured as follows, first we present our methods for producing neuromorphic LMs, using a copper coating and a CNT-containing solution core. After presenting our results, which detail an electrical characterisation of our LMs, including descriptions of  entrainment protocols, we proceed to discuss their design, putative uses and present limitations.

\section{Materials and Methods}

\begin{figure}[htbp]
    \centering
    \subfigure[]{\includegraphics[width=0.49\textwidth]{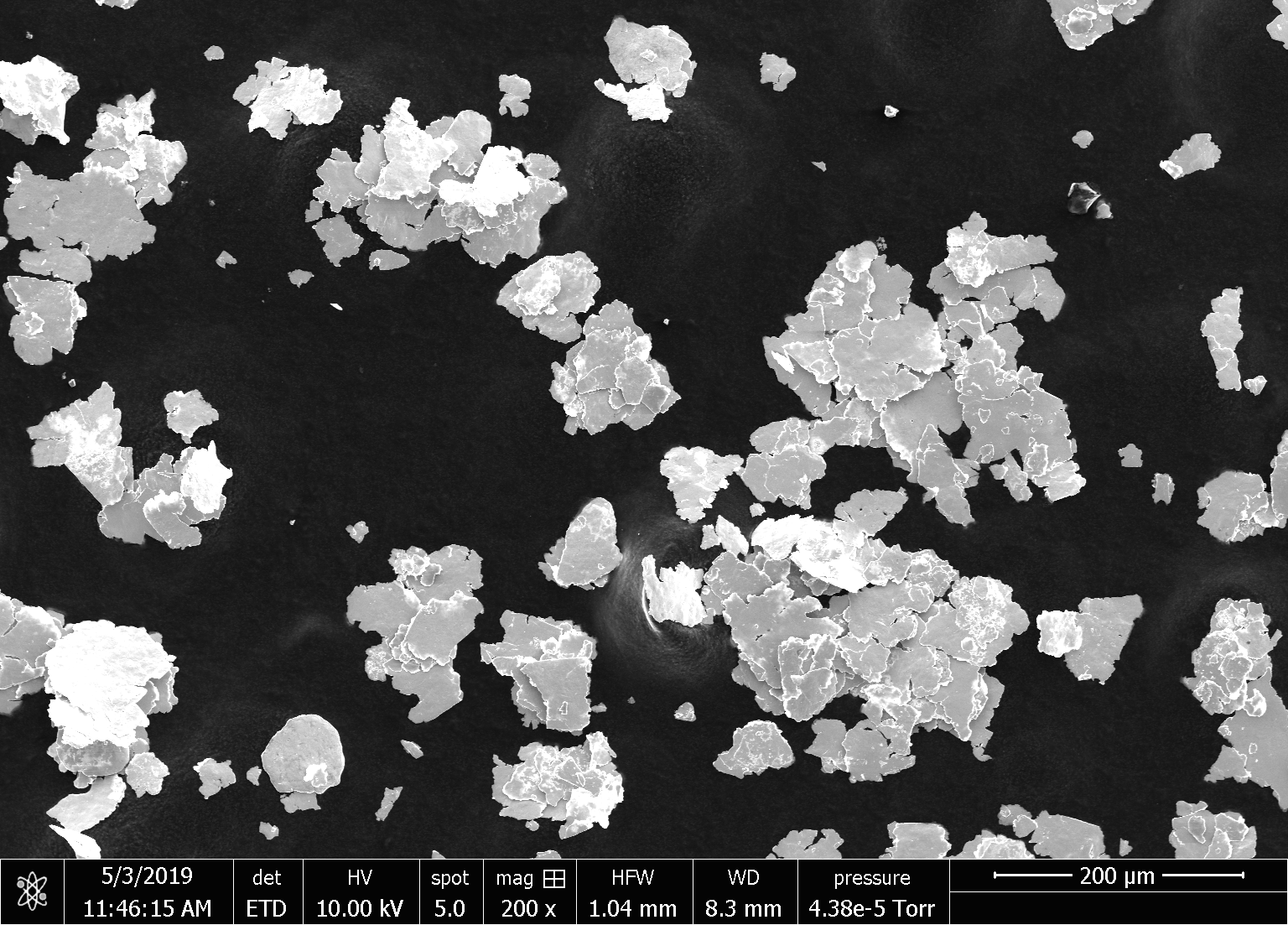}}
    \subfigure[]{\includegraphics[width=0.49\textwidth]{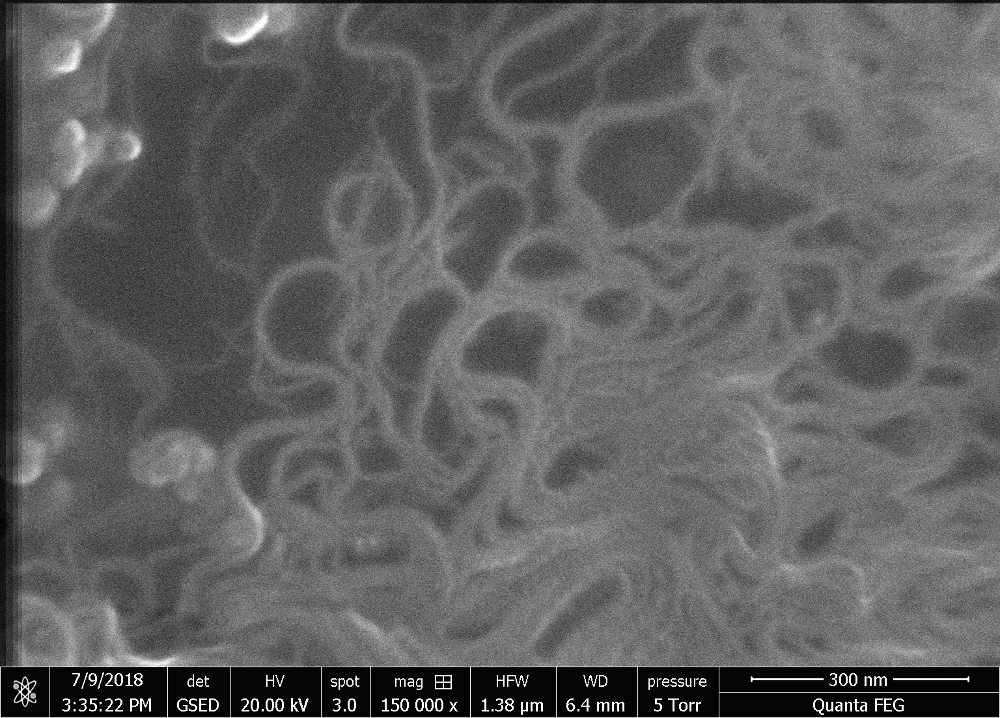}}
    \caption{Scanning electron micrographs of the experimental materials used. (a) Copper flakes, which were used for LM coatings. (b) Carbon nanotubes, used in LM cores.}
    \label{fig-sem}
\end{figure}

LMs were prepared using copper flakes (Goodfellow, UK) (average diameter \SI{76}{\micro\meter}, $n=838$) (Fig.~\ref{fig-sem}a) and \SI{100}{\micro\liter} liquid droplets. Upon gentle contact of the liquid with the bed of copper flakes, spontaneous LM formation was observed; the copper flakes migrated around the droplet of liquid, forming a completely coated LM without the need of rolling (this phenomenon has been previously observed using ethanol/water binary solutions and hydrophobised glass beads~\cite{Whitby2012}). The experimental LM fluid core was a single walled-CNT (Fig.~\ref{fig-sem}b) dispersion at \SI{1}{\milli\gram\per\milli\liter}, in deionised water (DIW) containing $1\,\%$~(w/v) Triton X-100 (Chasm Advanced Materials, USA). CNT dimensions were approximately \SI{20}{\nano\meter} diameter and over \SI{1}{\micro\meter} length. CNT solutions were sonicated prior to use for \SI{10}{\minute}. Control LMs were prepared using DIW (\SI{15}{\mega\ohm\centi\metre}) cores. Attempts were made to fabricate control LMs using 1\,\% Triton X-100 in DIW, but the surfactant nature of the additive in the absence of hydrophobic CNTs prevented LMs from forming. Previous reports using Triton X-100 (even with a more hydrophobic powder coating) also had limited success forming stable LMs at this concentration~\cite{Wang2018}. Care was taken to ascertain, both before and after experiments, that LMs had not wetted either electrode surface.

\begin{figure}[!htbp]
    \centering
    \subfigure[]{\includegraphics[width=0.49\textwidth]{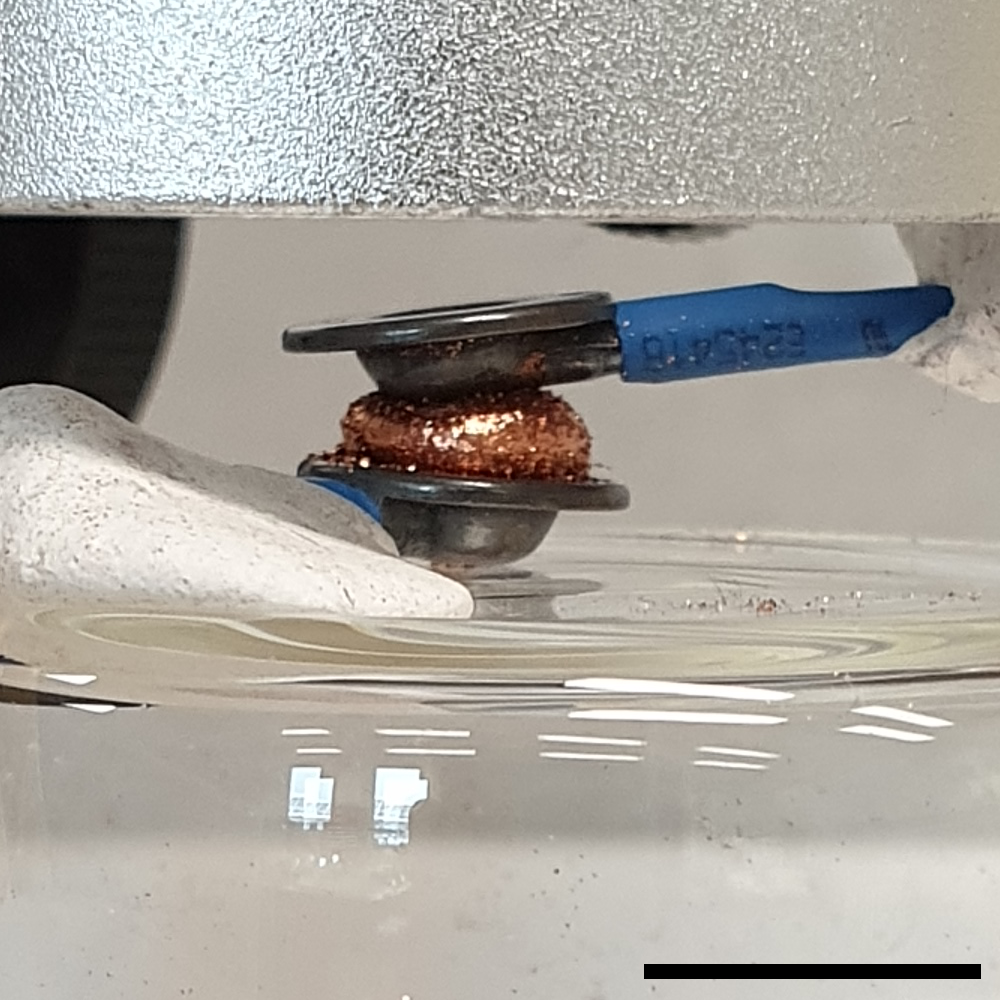}}
    \subfigure[]{\includegraphics[width=0.49\textwidth]{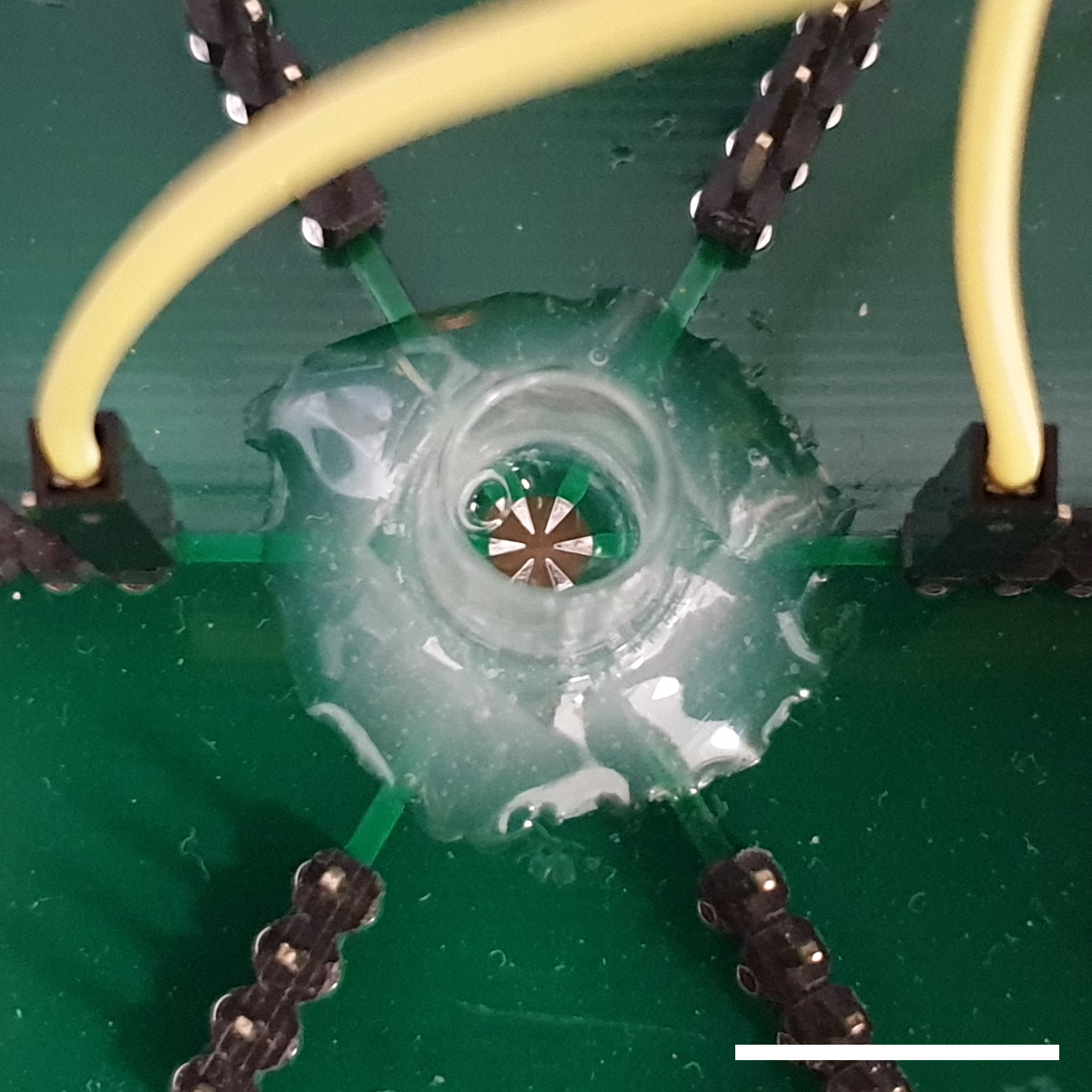}}
    \caption{Photographs of experimental electrical recording apparatus. (a) LM recording apparatus. Two 10~mm$^\slashcirc$ Ag/AgCl cup electrodes were used to `sandwich' LMs, with an electrode spacing of 2.0--2.5~mm. (b) Fluids were placed into wells overlying needle-shaped electrodes; only one pair of electrodes were used in the experiments described here. Scale bars 10~mm.}
    \label{fig_experiment}
\end{figure}

LMs were subject to electrical characterisation via two cup-style Ag/AgCl electrodes (Fig. \ref{fig_experiment}a), each with a total end-to-end resistance not exceeding \SI{0.5}{\ohm}. One electrode was placed concavity-side up and the LMs were rolled into the depression. The second electrode, mounted to a clamp stand, was then lowered onto the LM. The electrode spacing was assessed by eye to compress the LM slightly, at an inter-electrode gap of \SIrange{2.0}{2.5}{\milli\meter}, which ensured that the upper surface of the LMs remained in contact with the upper electrode in the event of LM deformation. These electrodes were chosen for their shape rather than their electrical properties, hence a brief series of replicate experiments were performed using flat copper plates instead of the cup electrodes, each measuring \SI[product-units = single]{10 x 10 x 1.0}{\milli\metre}, to ascertain whether the results observed were a function of the electrode material rather than the sample.

No steps were undertaken to prevent fluid evaporation from the LM because a)~a focus of our investigation was evolution of LM characteristics over time in standard room temperature and humidity conditions, and b)~our previous work \cite{Fullarton2018} has demonstrated that the effects of evaporation over the experiment duration (\SI{50}{\minute}) is likely to have been negligible.

Electrical measurements were made via a Keithley Source Measure Unit 2450 (Keithley Instruments, USA), using a 4-to-2 electrode setup. A current limit of \SI{100}{\milli\ampere} was used throughout. The principle experiment involved repeatedly stimulating the LM using a \SI{3.0}{\volt} pulse (\SI{0.5}{\second} ramp time, hence \SI{1.0}{\second} per pulse), followed by a delay (i.e.\ \SI{0}{\volt}) of the same duration. This stimulation pattern of $\SI{0}{\volt} \rightarrow \SI{3}{\volt}$ was repeated 375 times per `phase', each phase therefore lasting \SI{750}{\second}. Following, a second stimulation phase was started which was identical to the first, except for polarity being switched to \SI{-3.0}{\volt}. This overall pattern was repeated, i.e.\ $375$ pulses at \SI{3.0}{\volt}, then 375 pulses at \SI{-3.0}{\volt}, twice, such that each experiment's duration was \SI{3000}{\second}. These four phases are here named $s_1 \rightarrow s_4$. All experiments were repeated 10 times. Pulse frequency, duration and magnitude were all chosen as the result of prior testing, towards designing experiments where minimal voltages were used in order to reduce breakdown of water products whilst keeping experiments short enough to reduce the impact of fluid evaporation from marbles. For completeness, IV sweeps were also conducted with the same instrument using a \SI{3}{\volt} double-sided sweep, \SI{0.1}{\second} dwell time with a \SI{0.1}{\ampere} current limit.

Further control measurements were also collected on \SI{100}{\micro\liter} samples of fluids (dispersed CNTs and solvated Triton X-100 in DIW, Triton X-100 in DIW, both at the same concentrations as previously stated, and pure DIW), henceforth referred to as `free liquid' experiments. This was achieved through the use of bespoke circuit boards developed in our laboratory for bulk electrical testing of fluid samples (Fig.~\ref{fig_experiment}b), connected to the same measurement apparatus described above. Full details of these boards' design and fabrication are included in the ESI.

All analyses were performed using MatLab 2017a (MathWorks, USA). Scanning electron micrographs were recorded using a Quanta 650 field emission gun-scanning electron microscope (FEG-SEM) (FEI, USA).

\section{Results}
\subsection{Description of neuromorphic property}

An emergent characteristic was observed in LMs filled with CNT solution that will henceforth be known as a `neuromorphic switching effect' (NSE), the characteristics of which were as follows. Repeated stimulation with \SI{3}{\volt} pulses during $s_1$ caused CNT-filled LMs to maintain a high-resistance state (quantitative data are presented in subsection~\ref{sect-quanti}), during which fluctuations in their electrical properties were minor. The switching of polarity during $s_2$ resulted in the LM initially assuming a similar profile as during $s_1$, before suddenly switching to a more conductive profile. The majority of LMs experienced drops in resistance of one to two orders of magnitude during $s_2$. Entering phase $s_3$ (reverting to the original polarity), the LMs were observed to briefly retain their lower resistance profile before rapidly dropping back to a high resistance profile, similar to those observed in $s_1$. After a final polarity change during $s_4$, the LMs were observed to return to their low resistance profile. The LMs invariably switched to their low resistance profile in a shorter time period than during $s_2$.

\begin{figure}[htbp]
    \centering
    \includegraphics[width=\textwidth]{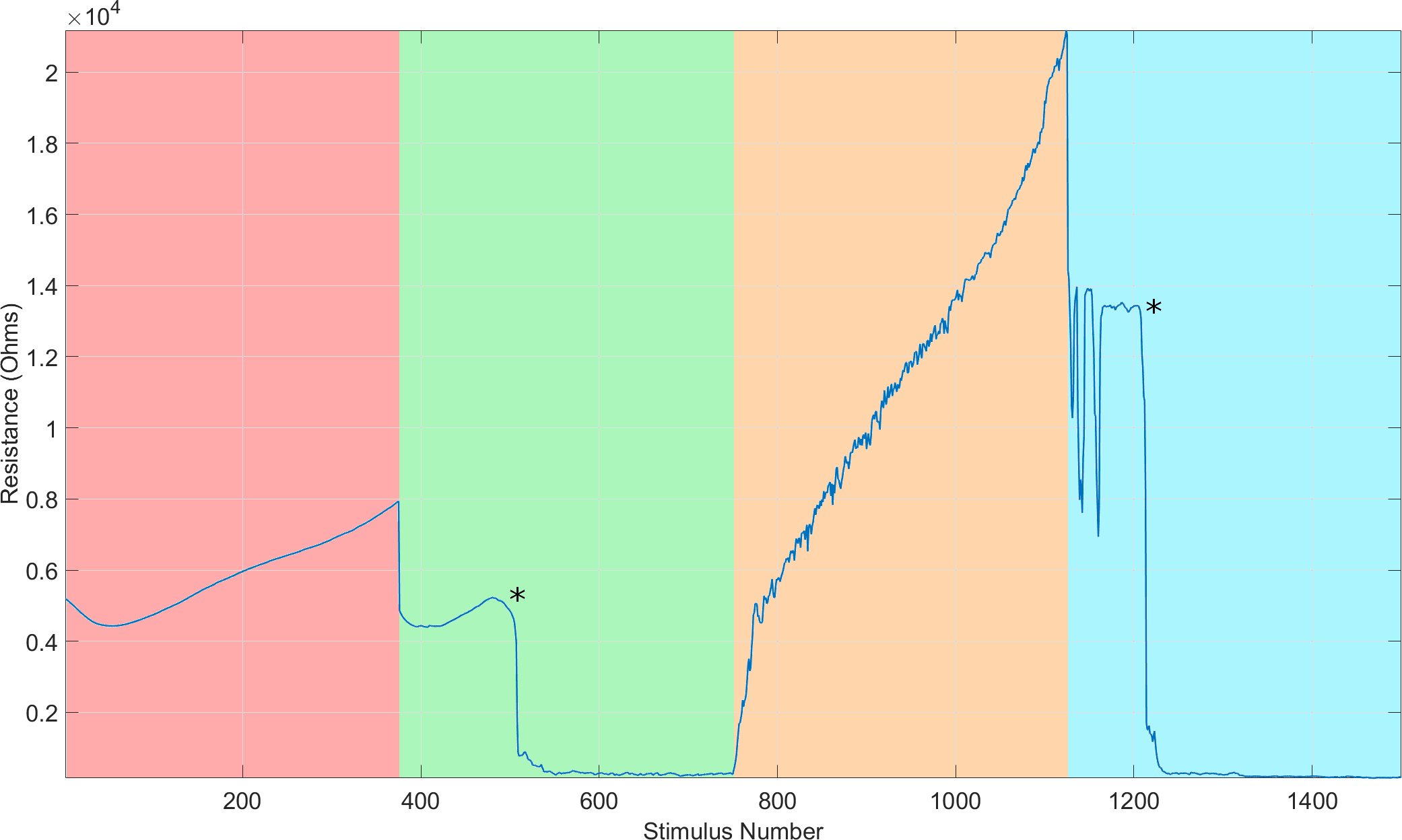}
    \caption{Graph to illustrate typical behaviour of a CNT-filled LM exposed to repeated stimulation with \SI{3}{\volt} at periodically alternating polarities, showing NSEs. Asterisks indicate onset of NSEs and coloured areas indicate stimulation phases (red for $s_1$ etc.). Resistance values at \SI{0}{\V} have been omitted for clarity.}
    \label{fig-exampleGraph}
\end{figure}

It is noteworthy that an NSE was not observed in water-filled control LMs, or in any of the free liquid experiments. The NSE was observed in CNT-filled LMs sandwiched between copper (rather than Ag/AgCl) electrodes, but not in water-filled LMs or any variety of free liquid when tested with copper electrodes. For example datasets, see ESI.

\subsection{Characterisation of neuromorphic system effect}
\label{sect-quanti}

All datasets were found to be normally distributed via Shapiro-Wilk tests (p $< 0.01$ in all instances). Differences in mean resistance of CNT-filled LMs between phases $s_1 \rightarrow s_4$ were found to be significant through Analysis of Variance (ANOVA) (Table \ref{table-anova}). There was also a significant difference in mean resistances of CNTs in free liquid between these phases (despite their resistance profile not exhibiting the NSE), but no significant difference between means between phases was observed in any other sample type. The resistance of CNT LMs was subject to a large degree of variation (which ranged from \SIrange[product-units = single]{5.67}{46.3}{\kilo\ohm} (mean  \SI{46.6}{\kilo\ohm}) in unstimulated $s_1$ CNT LMs and \SIrange[product-units = single]{0.05}{19.3}{\kilo\ohm} (mean \SI{3.15}{\kilo\ohm}) in $s_4$ CNT LMs), likely resulting from inconsistent contact areas between LMs and their top electrodes, hence data on conductivity changes are here presented as percentage changes in resistance between phases (Table \ref{table-ttest1}). Percentage differences in mean resistance between phases $s_1 \rightarrow s_2$ and $s_1 \rightarrow s_4$ were found to be significant, with variation reducing markedly in the latter measurement. 

This raises the question as to whether CNT LM resistance drops further in $s_4$ than in $s_2$. Whilst approximately 60\% of CNT LMs demonstrated a subsequent reduction in resistance $s_4$ from the values they exhibited during $s_2$, the remainder either continued to reproduce similar resistances or a slightly increased resistance, hence a significant difference in percentage change between $s_2 \rightarrow s_4$ was not observed.

\begin{table}[htbp]
\centering
\caption{Table to show ANOVA results, comparing differences in means between phases in LMs containing CNTs and water, in addition to free liquid (FL) controls containing CNTs, Triton X-100 (T) and water. $^\dagger$:~p~$< 0.05$.}
\begin{tabular}{c|c|c|c|c|c}
     & CNT LM & \ce{H2O} LM & CNT FL & T FL & \ce{H2O} FL \\ 
\hline
\hline 
F   &  3.236  & 1.086    & 3.490  & 0.700 & 1.970     \\
\hline
p    & 0.035$^\dagger$  & 0.392     & 0.041$^\dagger$  & 0.561 & 0.137    \\ 
\end{tabular} 
\label{table-anova}
\end{table}

\begin{table}[htbp]
\centering
\caption{Table to show one-sample t-test results, comparing percentage changes in resistance between $s_1 \rightarrow s_2$, $s_1 \rightarrow s_4$ and $s_2 \rightarrow s_4$ in LMs containing CNTs. CI: confidence interval (in \%), Med: median (in \%), PC: percentage change (in \%), SD: standard deviation, $^\dagger$:~p~$< 0.05$, $^\ddagger$:~p~$< 0.0001$.
}
\begin{tabular}{c|c|c|c|c|c|c}
                        & Mean PC & Med PC & CI Low & CI High & SD    & p  \\ 
\hline
\hline 
$s_1 \rightarrow s_2$   & 63.95    & 97.64 & 22.18  & 105.7   & 58.40  & 0.007$^\dagger$ \\
\hline
$s_1 \rightarrow s_4$   & 87.70    & 98.48 & 71.58  & 103.8   & 22.53  & 0.000$^\ddagger$ \\
\hline
$s_2 \rightarrow s_4$   & 29.23    & 38.67 & -13.82 & 72.27  & 60.17  & 0.159 \\
\end{tabular} 
\label{table-ttest1}
\end{table}

The time to NSE onset during phases $s_2$ and $s_4$ was also investigated. Similarly to the CNT LM's resistance, onset times were highly variable, ranging from \SIrange{11}{309}{\second}, mean \SI{84.8}{\second} during $s_2$ and \SIrange{4}{106}{\second}, mean \SI{41.8}{\second} during $s_4$. It was observed that longer time to NSE onset correlated with datasets with a higher overall resistance, hence this phenomenon was reasoned to also be linked to LM-electrode contact area. When measured as a percentage difference in onset between $s_2$ and $s_4$, it was found that the difference between time to onset was significantly shorter between $s_2$ and $s_4$ (Table \ref{table-ttest2}). 

\begin{table}[htbp]
\centering
\caption{Table to show two-sample t-test results, comparing percentage changes in mean NSE onset time between phases $s_2$ and $s_4$ in LMs containing CNTs. CI: confidence interval (in \%), Med: median (in \%), PC: percentage change (in \%), SD: standard deviation, $^*$:~p~$< 0.01$.}
\begin{tabular}{c|c|c|c|c|c|c}
                        & CI Low & CI High & Mean PC & Med PC & SD    & p \\
\hline
\hline 
$s_2 \rightarrow s_4$   & 25.71  & 60.69   & 43.20   & 39.73  & 22.75 & 0.0005$^*$ \\
\end{tabular} 
\label{table-ttest2}
\end{table}

\subsection{CNT LM IV profile}
\begin{figure}[htbp]
    \centering
    \includegraphics[width=\textwidth]{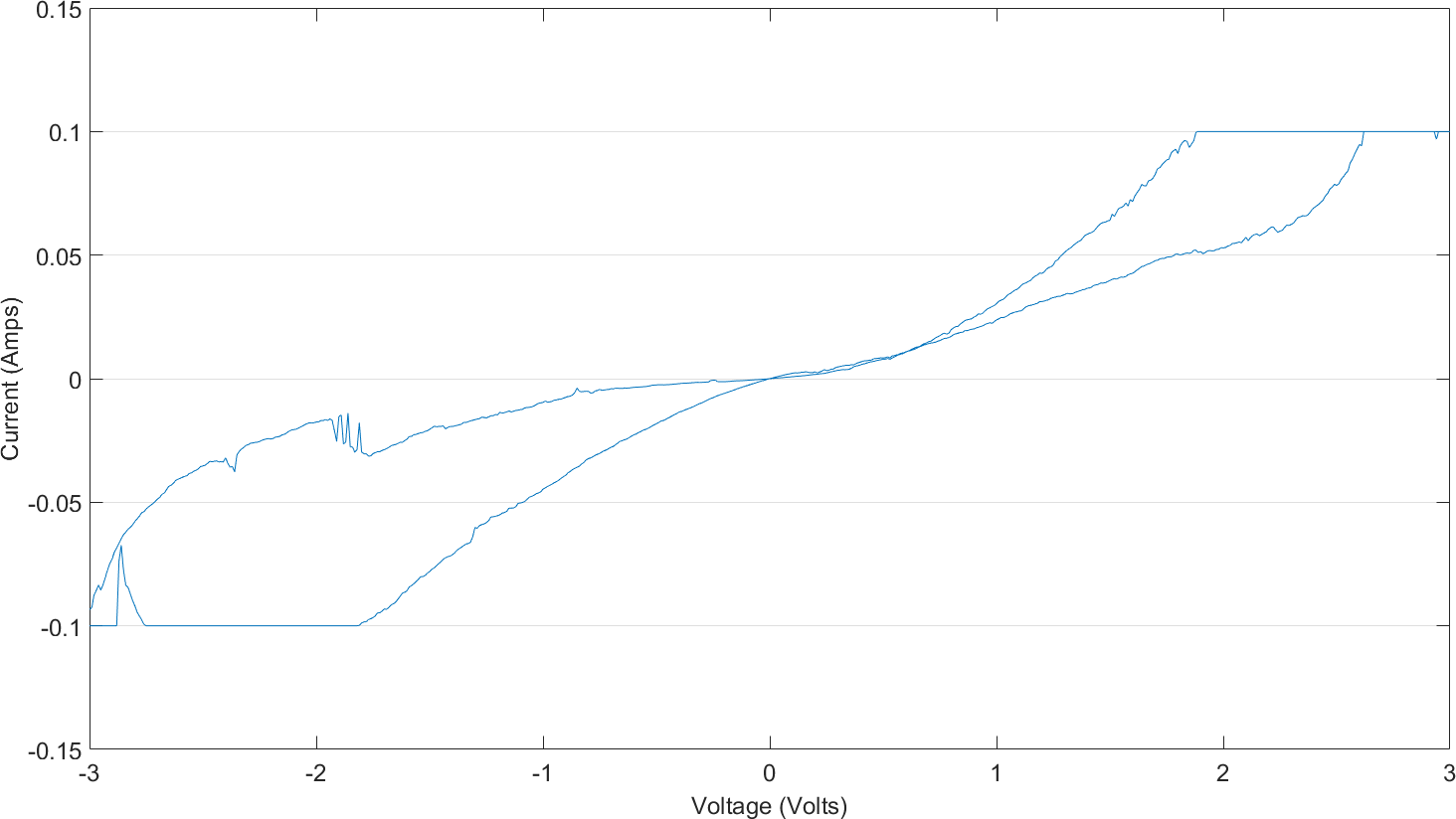}
    \caption{Graph to show a typical IV sweep profile from a CNT LM exposed to a \SI{3}{\volt} double-ended sweep. }
    \label{fig-iv}
\end{figure}

A typical IV profile for a previously-unstimulated CNT LM is shown in Fig.~\ref{fig-iv}. Allowing for the LMs saturating at the instrument current limit when exposed to a continuous DC source, all samples ($n=5$) were observed to produce a pinched-loop hysteresis, consistent with the description of a memristive device \cite{Biolek2014}. 

\section{Discussion}
We have described here laboratory experimental work in which we generated LMs which exhibit the neuromorphic properties of: (1)~Switching between two distinct electrochemical states in response to excitatory or inhibitory electrical input signals; (2)~Potentiation, i.e.~an increase in signal in response to repeated DC pulses (`training'); and (3)~Memory of previous states, as evidenced by a reduced time to NSE and IV profiles consistent with descriptions of memristors. We propose, therefore, that CNT LMs may be considered as soft non-biological synapses. Whilst CNTs have been successfully used as components of LM cores before~\cite{Nakai2013}, to our knowledge, this is the first published description of a neuromorphic LM.   

We did not observe any incidence of the NSE in CNT free liquid experiments, although as there was a statistically significant difference in mean resistance between phases, we are not discounting the possibility that this may still occur with alternate testing parameters. Regardless of whether this effect is a function purely of CNTs or otherwise, the NSE has specific applications when packaged within a LM. 

Whilst the mechanisms underlying the NSE in CNT LMs were not elucidated with the experiments detailed here, significant work has been done on the electrical properties of CNTs and their propensity to align according to electromagnetic fields is well-established~\cite{Sun2012}. Further, many neuromorphic devices based on CNT technologies have been proposed and work has begun on implanting CNTs into biological neurons, the effects of which appear to include neuromodulation~\cite{cellot2009carbon}.

\begin{figure}[htbp]
    \centering
    \subfigure[]{\includegraphics[width=0.37\textwidth]{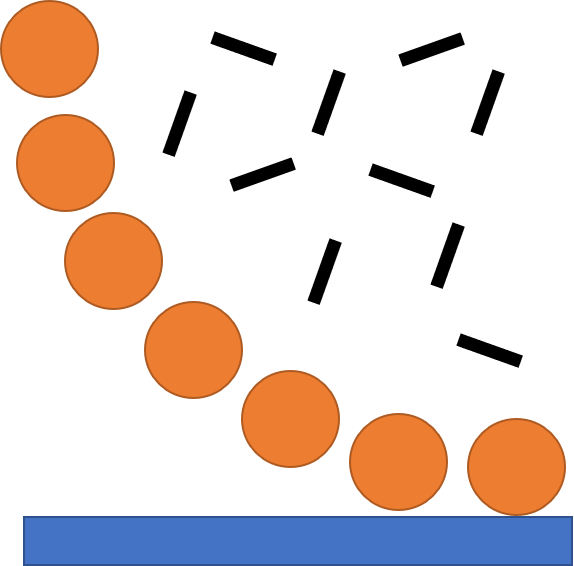}}
    \subfigure[]{\includegraphics[width=0.49\textwidth]{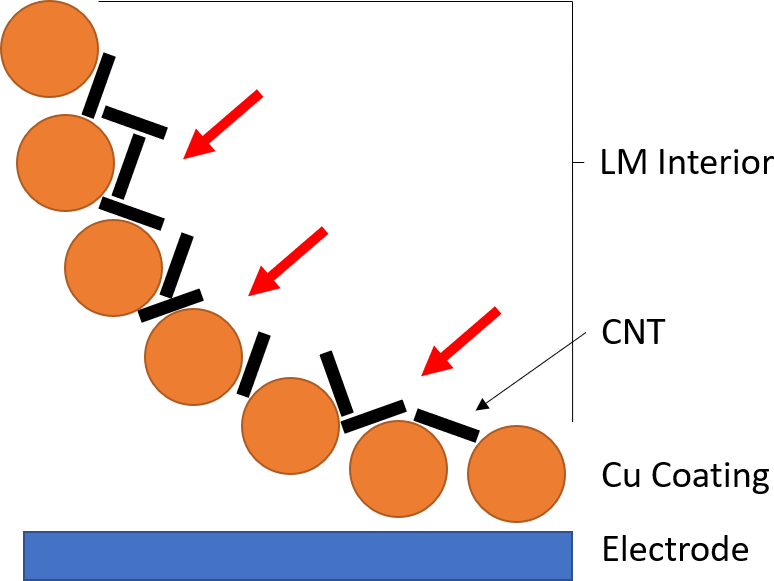}}
    \caption{Diagram to illustrate the hypothesised mechanics for NSE in CNT-filled LMs, show a 2D quarter-circle cross section (not to scale). (a) System during phases $s_1$ and $s_3$. (b) System during phases $s_2$ and $s_4$. Polarity reversal has caused CNT to associate with the LM copper coating, creating pathways of increased conductivity. Large red arrows indicate direction of CNT movement in response to polarity switching.}
    \label{fig-hx}
\end{figure}

We hypothesise that the mechanism underlying the NSE in our CNT LMs involves charging of CNTs in response to repeated electrical stimulation~\cite{poncharal1999electrostatic,maciel2008electron,matsunaga2011observation}, which results in their being attracted to the metallic coating of the LM under one polarity of DC stimulation and repelled when the polarity is reversed. The effect of this would be to induce extra electrical connections to be made between the particles of metal coating, thereby reducing the LM's resistance (Fig.~\ref{fig-hx}). This is somewhat consistent with results on CNTs sustaining and promoting neuronal electrical activity in networks of cultured cells as reported by Cellot \textit{et al.}~\cite{cellot2009carbon}; where it is proposed that the reported effects could be due to clustering of CNTs and the resulting close contact with the neural membrane. 

Switching behaviour may be used as the basis for computing circuitry, including logic gates and bistables: this would be the most facile route towards producing CNT LM computing devices. When considering their neuromorphic characteristics, however, even a single memristive device which exhibits spiking behaviour may be capable of implementing combinatorial logic operations (e.g.\ full adders) when device I/O operations are sequence-sensitive \cite{Gale2019,Diedrich2018}. It is therefore clear that more refined methods are available to enhance the viability of CNT LMs as unconventional computing media. 

The benefits to our devices being encapsulated within liquid marbles are manifold~\cite{McHale2011,McHale2015}, but revolve around their representing soft, ballistic data sources whose contents may be considered as chemical reactors. Through the exploitation of principles of collision-based computing \cite{adamatzky2002cbc}, LM computing devices may be used to implement non-standard, `collision-based' conservative logics~\cite{draper2017liquid}. Integration of LM features such as collisions whose outcome may be engineered (reflection or coalescence), and the potential for chemical reactions between two heterogenous fluid cores following collision, further enhances the toolbox of traditional conservative logic. The applications of such a device range from enhancing our understanding of the nervous system and hence information processing in mammals, in addition to various lab-on-a-chip applications.

Our future work on neuromorphic LMs will establish the number of stimuli and switching cycles necessary to produce an NSE, their longevity under these conditions and how consistent memory effects are over time. Further work on elucidating the mechanism underlying the phenomena described here will also be prioritised.


\section{Conclusions}
Key characteristics of CNT LMs are as follows.

\begin{enumerate}
    \item Sensitization of the LM and/or their contents causes a rapid drop in resistance shortly after inverting the polarity of stimulation (the NSE).
    \item Repeated stimulation across multiple phases causes a more reliable and more stable decrease in LM resistance, which may be equated with the concept of training.
    \item Repeated stimulation across multiple phases causes the NSE to occur more rapidly, which may also be equated with training.
\end{enumerate}

We propose that this technology is of interest to the design and fabrication of massively-parallel wet computers whose applications range from computing to biomedicine.

\section*{Acknowledgements}
The authors would like to extend their gratitude to Dr David Patton and Mrs Sue Hula, of the Department of Applied Sciences at UWE Bristol, for their expertise with the scanning electron microscope. TCD, CF, BPJDLC and AA acknowledge support of \mbox{EPSRC} with grant \mbox{EP/P016677/1}. 

\section*{Declaration of Interest}
The authors declare no competing interests.

\section*{Key to ESI}
\noindent
Section 1: Description of hex board design and construction\\
Section 2: Example graphs representative of control measurements\\

\section*{Data Availability Statement}
The raw/processed data required to reproduce these findings cannot be shared at this time due to technical or time limitations. Data will be made available on request.

\section*{References}
\bibliographystyle{model1-num-names}
\bibliography{references.bib}

\begin{thebibliography}{49}
\expandafter\ifx\csname natexlab\endcsname\relax\def\natexlab#1{#1}\fi
\providecommand{\bibinfo}[2]{#2}
\ifx\xfnm\relax \def\xfnm[#1]{\unskip,\space#1}\fi
\bibitem[{Priel et~al.(2010)Priel, Tuszynski, and Woolf}]{Priel2010}
\bibinfo{author}{A.~Priel}, \bibinfo{author}{J.~a. Tuszynski},
  \bibinfo{author}{N.~J. Woolf},
\newblock \bibinfo{title}{{Neural cytoskeleton capabilities for learning and
  memory.}},
\newblock \bibinfo{journal}{Journal of biological physics} \bibinfo{volume}{36}
  (\bibinfo{year}{2010}) \bibinfo{pages}{3--21}.
\bibitem[{Mayne et~al.(2015)Mayne, Adamatzky, and Jones}]{Mayne2014cib}
\bibinfo{author}{R.~Mayne}, \bibinfo{author}{A.~Adamatzky},
  \bibinfo{author}{J.~Jones},
\newblock \bibinfo{title}{{On the role of the plasmodial cytoskeleton in
  facilitating intelligent behaviour in slime mould Physarum polycephalum}},
\newblock \bibinfo{journal}{Communicative and Integrative Biology}
  \bibinfo{volume}{8} (\bibinfo{year}{2015}) \bibinfo{pages}{e1059007}.
\bibitem[{Adamatzky and Mayne(2015)}]{Adamatzky2015actinautomata}
\bibinfo{author}{A.~Adamatzky}, \bibinfo{author}{R.~Mayne},
\newblock \bibinfo{title}{{Actin Automata: Phenomenology and Localizations}},
\newblock \bibinfo{journal}{International Journal of Bifurcation and Chaos}
  \bibinfo{volume}{25} (\bibinfo{year}{2015}).
\bibitem[{Mayne and Adamatzky(2014)}]{Mayne2014}
\bibinfo{author}{R.~Mayne}, \bibinfo{author}{A.~Adamatzky},
\newblock \bibinfo{title}{{The Physarum polycephalum actin network:
  formalisation , topology and morphological correlates with computational
  ability}},
\newblock in: \bibinfo{booktitle}{8th International Conference on Bio-inspired
  Information and Communications Technologies (formerly BIONETICS)}, pp.
  \bibinfo{pages}{87--94}.
\bibitem[{Woolf et~al.(2010)Woolf, Priel, and Tuszy{\'{n}}ski}]{Woolf2010}
\bibinfo{author}{N.~Woolf}, \bibinfo{author}{A.~Priel},
  \bibinfo{author}{J.~Tuszy{\'{n}}ski},
\newblock \bibinfo{title}{{The cytoskeleton as a nanoscale information
  processor: electrical properties and an actin-microtubule network model}},
\newblock in: \bibinfo{editor}{N.~Woolf}, \bibinfo{editor}{A.~Priel},
  \bibinfo{editor}{J.~Tuszynski} (Eds.), \bibinfo{booktitle}{Nanoneuroscience.
  Structural and Functional Roles of the Neuronal Cytoskeleton in Health and
  Disease}, \bibinfo{publisher}{Springer}, \bibinfo{address}{Berlin},
  \bibinfo{year}{2010}, pp. \bibinfo{pages}{85--127}.
\bibitem[{Hameroff(1987)}]{Hameroff1987}
\bibinfo{author}{S.~R. Hameroff}, \bibinfo{title}{{Ultimate computing}},
  \bibinfo{publisher}{North-Holland}, \bibinfo{address}{Amsterdam},
  \bibinfo{year}{1987}.
\bibitem[{Lahoz-Beltra et~al.(1993)Lahoz-Beltra, Hameroff, and
  Dayhoff}]{Lahoz-Beltra1993}
\bibinfo{author}{R.~Lahoz-Beltra}, \bibinfo{author}{S.~Hameroff},
  \bibinfo{author}{J.~Dayhoff},
\newblock \bibinfo{title}{{Cytoskeletal logic: a model for molecular
  computation via Boolean operations in microtubules and microtubule-associated
  proteins}},
\newblock \bibinfo{journal}{Biosystems} \bibinfo{volume}{29}
  (\bibinfo{year}{1993}) \bibinfo{pages}{1--23}.
\bibitem[{Whiting et~al.(2016)Whiting, Jones, Bull, Levin, and
  Adamatzky}]{Whiting2016}
\bibinfo{author}{J.~Whiting}, \bibinfo{author}{J.~Jones},
  \bibinfo{author}{L.~Bull}, \bibinfo{author}{M.~Levin},
  \bibinfo{author}{A.~Adamatzky},
\newblock \bibinfo{title}{Towards a physarum learning chip},
\newblock \bibinfo{journal}{Scientific Reports} \bibinfo{volume}{19948}
  (\bibinfo{year}{2016}).
\bibitem[{Aussillous and Qu{\'e}r{\'e}(2001)}]{aussillous2001liquid}
\bibinfo{author}{P.~Aussillous}, \bibinfo{author}{D.~Qu{\'e}r{\'e}},
\newblock \bibinfo{title}{Liquid marbles},
\newblock \bibinfo{journal}{Nature} \bibinfo{volume}{411}
  (\bibinfo{year}{2001}) \bibinfo{pages}{924}.
\bibitem[{McHale and Newton(2011)}]{McHale2011}
\bibinfo{author}{G.~McHale}, \bibinfo{author}{M.~I. Newton},
\newblock \bibinfo{title}{{Liquid marbles: principles and applications}},
\newblock \bibinfo{journal}{Soft Matter} \bibinfo{volume}{7}
  (\bibinfo{year}{2011}) \bibinfo{pages}{5473}.
\bibitem[{McHale and Newton(2015)}]{McHale2015}
\bibinfo{author}{G.~McHale}, \bibinfo{author}{M.~I. Newton},
\newblock \bibinfo{title}{{Liquid marbles: topical context within soft matter
  and recent progress}},
\newblock \bibinfo{journal}{Soft Matter} \bibinfo{volume}{11}
  (\bibinfo{year}{2015}) \bibinfo{pages}{2530--2546}.
\bibitem[{Bormashenko(2016)}]{bormashenko2016liquid}
\bibinfo{author}{E.~Bormashenko},
\newblock \bibinfo{title}{Liquid marbles, elastic nonstick droplets: From
  minireactors to self-propulsion},
\newblock \bibinfo{journal}{Langmuir} \bibinfo{volume}{33}
  (\bibinfo{year}{2016}) \bibinfo{pages}{663--669}.
\bibitem[{Aussillous and Qu{\'e}r{\'e}(2006)}]{aussillous2006properties}
\bibinfo{author}{P.~Aussillous}, \bibinfo{author}{D.~Qu{\'e}r{\'e}},
\newblock \bibinfo{title}{Properties of liquid marbles},
\newblock \bibinfo{journal}{Proceedings of the Royal Society A: Mathematical,
  Physical and Engineering Sciences} \bibinfo{volume}{462}
  (\bibinfo{year}{2006}) \bibinfo{pages}{973--999}.
\bibitem[{Bormashenko et~al.(2009)Bormashenko, Bormashenko, Musin, and
  Barkay}]{bormashenko2009mechanism}
\bibinfo{author}{E.~Bormashenko}, \bibinfo{author}{Y.~Bormashenko},
  \bibinfo{author}{A.~Musin}, \bibinfo{author}{Z.~Barkay},
\newblock \bibinfo{title}{On the mechanism of floating and sliding of liquid
  marbles},
\newblock \bibinfo{journal}{ChemPhysChem} \bibinfo{volume}{10}
  (\bibinfo{year}{2009}) \bibinfo{pages}{654--656}.
\bibitem[{Bormashenko et~al.(2010)Bormashenko, Bormashenko, and
  Oleg}]{bormashenko2010nature}
\bibinfo{author}{E.~Bormashenko}, \bibinfo{author}{Y.~Bormashenko},
  \bibinfo{author}{G.~Oleg},
\newblock \bibinfo{title}{On the nature of the friction between nonstick
  droplets and solid substrates},
\newblock \bibinfo{journal}{Langmuir} \bibinfo{volume}{26}
  (\bibinfo{year}{2010}) \bibinfo{pages}{12479--12482}.
\bibitem[{Pike et~al.(2002)Pike, Richard, Foster, and Mahadevan}]{Pike2002}
\bibinfo{author}{N.~Pike}, \bibinfo{author}{D.~Richard},
  \bibinfo{author}{W.~Foster}, \bibinfo{author}{L.~Mahadevan},
\newblock \bibinfo{title}{{How aphids lose their marbles}},
\newblock \bibinfo{journal}{Proc. R. Soc. B Biol. Sci.} \bibinfo{volume}{269}
  (\bibinfo{year}{2002}) \bibinfo{pages}{1211--1215}.
\bibitem[{Kasahara et~al.(2019)Kasahara, Akimoto, Hariyama, Takaku, Yusa,
  Okada, Nakajima, Hirai, Mayama, Okada, Deguchi, Nakamura, and
  Fujii}]{Kasahara2019}
\bibinfo{author}{M.~Kasahara}, \bibinfo{author}{S.-i. Akimoto},
  \bibinfo{author}{T.~Hariyama}, \bibinfo{author}{Y.~Takaku},
  \bibinfo{author}{S.-i. Yusa}, \bibinfo{author}{S.~Okada},
  \bibinfo{author}{K.~Nakajima}, \bibinfo{author}{T.~Hirai},
  \bibinfo{author}{H.~Mayama}, \bibinfo{author}{S.~Okada},
  \bibinfo{author}{S.~Deguchi}, \bibinfo{author}{Y.~Nakamura},
  \bibinfo{author}{S.~Fujii},
\newblock \bibinfo{title}{{Liquid Marbles in Nature: Craft of Aphids for
  Survival}},
\newblock \bibinfo{journal}{Langmuir} \bibinfo{volume}{35}
  (\bibinfo{year}{2019}) \bibinfo{pages}{6169--6178}.
\bibitem[{Bormashenko(2011)}]{bormashenko2011liquid}
\bibinfo{author}{E.~Bormashenko},
\newblock \bibinfo{title}{Liquid marbles: properties and applications},
\newblock \bibinfo{journal}{Current Opinion in Colloid \& Interface Science}
  \bibinfo{volume}{16} (\bibinfo{year}{2011}) \bibinfo{pages}{266--271}.
\bibitem[{Ooi and Nguyen(2015)}]{ooi2015manipulation}
\bibinfo{author}{C.~H. Ooi}, \bibinfo{author}{N.-T. Nguyen},
\newblock \bibinfo{title}{Manipulation of liquid marbles},
\newblock \bibinfo{journal}{Microfluidics and Nanofluidics}
  \bibinfo{volume}{19} (\bibinfo{year}{2015}) \bibinfo{pages}{483--495}.
\bibitem[{Oliveira et~al.(2017)Oliveira, Reis, and
  Mano}]{oliveira2017potential}
\bibinfo{author}{N.~M. Oliveira}, \bibinfo{author}{R.~L. Reis},
  \bibinfo{author}{J.~F. Mano},
\newblock \bibinfo{title}{The potential of liquid marbles for biomedical
  applications: A critical review},
\newblock \bibinfo{journal}{Advanced healthcare materials} \bibinfo{volume}{6}
  (\bibinfo{year}{2017}) \bibinfo{pages}{1700192}.
\bibitem[{Tian et~al.(2010{\natexlab{a}})Tian, Arbatan, Li, and
  Shen}]{Tian2010}
\bibinfo{author}{J.~Tian}, \bibinfo{author}{T.~Arbatan},
  \bibinfo{author}{X.~Li}, \bibinfo{author}{W.~Shen},
\newblock \bibinfo{title}{{Liquid marble for gas sensing}},
\newblock \bibinfo{journal}{Chem. Commun.} \bibinfo{volume}{46}
  (\bibinfo{year}{2010}{\natexlab{a}}) \bibinfo{pages}{4734}.
\bibitem[{Tian et~al.(2010{\natexlab{b}})Tian, Arbatan, Li, and
  Shen}]{Tian2010a}
\bibinfo{author}{J.~Tian}, \bibinfo{author}{T.~Arbatan},
  \bibinfo{author}{X.~Li}, \bibinfo{author}{W.~Shen},
\newblock \bibinfo{title}{{Porous liquid marble shell offers possibilities for
  gas detection and gas reactions}},
\newblock \bibinfo{journal}{Chem. Eng. J.} \bibinfo{volume}{165}
  (\bibinfo{year}{2010}{\natexlab{b}}) \bibinfo{pages}{347--353}.
\bibitem[{Draper et~al.(2017)Draper, Fullarton, Phillips, de~Lacy~Costello, and
  Adamatzky}]{draper2017liquid}
\bibinfo{author}{T.~C. Draper}, \bibinfo{author}{C.~Fullarton},
  \bibinfo{author}{N.~Phillips}, \bibinfo{author}{B.~P. de~Lacy~Costello},
  \bibinfo{author}{A.~Adamatzky},
\newblock \bibinfo{title}{Liquid marble interaction gate for collision-based
  computing},
\newblock \bibinfo{journal}{Materials Today} \bibinfo{volume}{20}
  (\bibinfo{year}{2017}) \bibinfo{pages}{561--568}.
\bibitem[{Draper et~al.(2018{\natexlab{a}})Draper, Fullarton, Phillips,
  de~Lacy~Costello, and Adamatzky}]{draper2018liquid}
\bibinfo{author}{T.~C. Draper}, \bibinfo{author}{C.~Fullarton},
  \bibinfo{author}{N.~Phillips}, \bibinfo{author}{B.~P. de~Lacy~Costello},
  \bibinfo{author}{A.~Adamatzky},
\newblock \bibinfo{title}{Liquid marble actuator for microfluidic logic
  systems},
\newblock \bibinfo{journal}{Scientific reports} \bibinfo{volume}{8}
  (\bibinfo{year}{2018}{\natexlab{a}}) \bibinfo{pages}{14153}.
\bibitem[{Draper et~al.(2018{\natexlab{b}})Draper, Fullarton, Phillips, {de
  Lacy Costello}, and Adamatzky}]{draper2018ucnc}
\bibinfo{author}{T.~C. Draper}, \bibinfo{author}{C.~Fullarton},
  \bibinfo{author}{N.~Phillips}, \bibinfo{author}{B.~P.~J. {de Lacy Costello}},
  \bibinfo{author}{A.~Adamatzky},
\newblock \bibinfo{title}{{Mechanical Sequential Counting with Liquid
  Marbles}},
\newblock in: \bibinfo{editor}{S.~Stepney}, \bibinfo{editor}{S.~Verlan} (Eds.),
  \bibinfo{booktitle}{UCNC 2018, LNCS 10867}, \bibinfo{publisher}{Springer},
  \bibinfo{year}{2018}{\natexlab{b}}, pp. \bibinfo{pages}{59--71}.
\bibitem[{Fullarton et~al.(2018)Fullarton, Draper, Phillips, Mayne, {De Lacy
  Costello}, and Adamatzky}]{Fullarton2018}
\bibinfo{author}{C.~Fullarton}, \bibinfo{author}{T.~Draper},
  \bibinfo{author}{N.~Phillips}, \bibinfo{author}{R.~Mayne},
  \bibinfo{author}{B.~{De Lacy Costello}}, \bibinfo{author}{A.~Adamatzky},
\newblock \bibinfo{title}{{Evaporation, Lifetime, and Robustness Studies of
  Liquid Marbles for Collision-Based Computing}},
\newblock \bibinfo{journal}{Langmuir} \bibinfo{volume}{34}
  (\bibinfo{year}{2018}).
\bibitem[{Dandan and Erbil(2009)}]{Dandan2009}
\bibinfo{author}{M.~Dandan}, \bibinfo{author}{H.~Y. Erbil},
\newblock \bibinfo{title}{{Evaporation Rate of Graphite Liquid Marbles:
  Comparison with Water Droplets}},
\newblock \bibinfo{journal}{Langmuir} \bibinfo{volume}{25}
  (\bibinfo{year}{2009}) \bibinfo{pages}{8362--8367}.
\bibitem[{Draper et~al.(2019)Draper, Fullarton, Mayne, Phillips, Canciani, {de
  Lacy Costello}, and Adamatzky}]{Draper2019collide}
\bibinfo{author}{T.~C. Draper}, \bibinfo{author}{C.~Fullarton},
  \bibinfo{author}{R.~Mayne}, \bibinfo{author}{N.~Phillips},
  \bibinfo{author}{G.~E. Canciani}, \bibinfo{author}{B.~P.~J. {de Lacy
  Costello}}, \bibinfo{author}{A.~Adamatzky},
\newblock \bibinfo{title}{{Mapping outcomes of liquid marble collisions}},
\newblock \bibinfo{journal}{Soft Matter} \bibinfo{volume}{15}
  (\bibinfo{year}{2019}) \bibinfo{pages}{3541--3551}.
\bibitem[{Planchette et~al.(2013)Planchette, Biance, Pitois, and
  Lorenceau}]{Planchette2013}
\bibinfo{author}{C.~Planchette}, \bibinfo{author}{A.-L. Biance},
  \bibinfo{author}{O.~Pitois}, \bibinfo{author}{E.~Lorenceau},
\newblock \bibinfo{title}{{Coalescence of armored interface under impact}},
\newblock \bibinfo{journal}{Phys. Fluids} \bibinfo{volume}{25}
  (\bibinfo{year}{2013}) \bibinfo{pages}{042104}.
\bibitem[{Cui et~al.(2002)Cui, Sordan, Burghard, and Kern}]{cui2002carbon}
\bibinfo{author}{J.~Cui}, \bibinfo{author}{R.~Sordan},
  \bibinfo{author}{M.~Burghard}, \bibinfo{author}{K.~Kern},
\newblock \bibinfo{title}{Carbon nanotube memory devices of high charge storage
  stability},
\newblock \bibinfo{journal}{Applied Physics Letters} \bibinfo{volume}{81}
  (\bibinfo{year}{2002}) \bibinfo{pages}{3260--3262}.
\bibitem[{Heinze et~al.(2002)Heinze, Tersoff, Martel, Derycke, Appenzeller, and
  Avouris}]{heinze2002carbon}
\bibinfo{author}{S.~Heinze}, \bibinfo{author}{J.~Tersoff},
  \bibinfo{author}{R.~Martel}, \bibinfo{author}{V.~Derycke},
  \bibinfo{author}{J.~Appenzeller}, \bibinfo{author}{P.~Avouris},
\newblock \bibinfo{title}{Carbon nanotubes as schottky barrier transistors},
\newblock \bibinfo{journal}{Physical Review Letters} \bibinfo{volume}{89}
  (\bibinfo{year}{2002}) \bibinfo{pages}{106801}.
\bibitem[{Bertin et~al.(2006)Bertin, Rueckes, and Segal}]{bertin2006nanotube}
\bibinfo{author}{C.~L. Bertin}, \bibinfo{author}{T.~Rueckes},
  \bibinfo{author}{B.~M. Segal}, \bibinfo{title}{Nanotube-based switching
  elements with multiple controls}, \bibinfo{year}{2006}. \bibinfo{note}{US
  Patent 6,990,009}.
\bibitem[{Choi et~al.(2006)Choi, Yoo, and Chu}]{choi2006memory}
\bibinfo{author}{W.-b. Choi}, \bibinfo{author}{I.-k. Yoo},
  \bibinfo{author}{J.-u. Chu}, \bibinfo{title}{Memory device utilizing carbon
  nanotubes}, \bibinfo{year}{2006}. \bibinfo{note}{US Patent 7,015,500}.
\bibitem[{Avouris(2002)}]{avouris2002molecular}
\bibinfo{author}{P.~Avouris},
\newblock \bibinfo{title}{Molecular electronics with carbon nanotubes},
\newblock \bibinfo{journal}{Accounts of chemical research} \bibinfo{volume}{35}
  (\bibinfo{year}{2002}) \bibinfo{pages}{1026--1034}.
\bibitem[{Diehl et~al.(2003)Diehl, Steuerman, Tseng, Vignon, Star, Celestre,
  Stoddart, and Heath}]{diehl2003single}
\bibinfo{author}{M.~R. Diehl}, \bibinfo{author}{D.~W. Steuerman},
  \bibinfo{author}{H.-R. Tseng}, \bibinfo{author}{S.~A. Vignon},
  \bibinfo{author}{A.~Star}, \bibinfo{author}{P.~C. Celestre},
  \bibinfo{author}{J.~F. Stoddart}, \bibinfo{author}{J.~R. Heath},
\newblock \bibinfo{title}{Single-walled carbon nanotube based molecular switch
  tunnel junctions},
\newblock \bibinfo{journal}{ChemPhysChem} \bibinfo{volume}{4}
  (\bibinfo{year}{2003}) \bibinfo{pages}{1335--1339}.
\bibitem[{Kim et~al.(2013)Kim, Chen, Truong, Shen, and Chen}]{kim2013carbon}
\bibinfo{author}{K.~Kim}, \bibinfo{author}{C.-L. Chen},
  \bibinfo{author}{Q.~Truong}, \bibinfo{author}{A.~M. Shen},
  \bibinfo{author}{Y.~Chen},
\newblock \bibinfo{title}{A carbon nanotube synapse with dynamic logic and
  learning},
\newblock \bibinfo{journal}{Advanced Materials} \bibinfo{volume}{25}
  (\bibinfo{year}{2013}) \bibinfo{pages}{1693--1698}.
\bibitem[{Kim et~al.(2017)Kim, Choi, Lim, Yoon, Lee, Kim, and
  Choi}]{kim2017pattern}
\bibinfo{author}{S.~Kim}, \bibinfo{author}{B.~Choi}, \bibinfo{author}{M.~Lim},
  \bibinfo{author}{J.~Yoon}, \bibinfo{author}{J.~Lee}, \bibinfo{author}{H.-D.
  Kim}, \bibinfo{author}{S.-J. Choi},
\newblock \bibinfo{title}{Pattern recognition using carbon nanotube synaptic
  transistors with an adjustable weight update protocol},
\newblock \bibinfo{journal}{ACS nano} \bibinfo{volume}{11}
  (\bibinfo{year}{2017}) \bibinfo{pages}{2814--2822}.
\bibitem[{Whitby et~al.(2012)Whitby, Bian, and Sedev}]{Whitby2012}
\bibinfo{author}{C.~P. Whitby}, \bibinfo{author}{X.~Bian},
  \bibinfo{author}{R.~Sedev},
\newblock \bibinfo{title}{{Spontaneous liquid marble formation on packed porous
  beds}},
\newblock \bibinfo{journal}{Soft Matter} \bibinfo{volume}{8}
  (\bibinfo{year}{2012}) \bibinfo{pages}{11336}.
\bibitem[{Wang and He(2018)}]{Wang2018}
\bibinfo{author}{C.~Wang}, \bibinfo{author}{Y.~He},
\newblock \bibinfo{title}{{Timed disintegrating of the liquid marbles
  containing triton X-100}},
\newblock \bibinfo{journal}{Colloids Surfaces A Physicochem. Eng. Asp.}
  \bibinfo{volume}{558} (\bibinfo{year}{2018}) \bibinfo{pages}{367--372}.
\bibitem[{{Biolek} et~al.(2014){Biolek}, {Biolkova}, and {Kolka}}]{Biolek2014}
\bibinfo{author}{D.~{Biolek}}, \bibinfo{author}{V.~{Biolkova}},
  \bibinfo{author}{Z.~{Kolka}},
\newblock \bibinfo{title}{Memristor pinched hysteresis loops: Touching points,
  part i},
\newblock in: \bibinfo{booktitle}{2014 International Conference on Applied
  Electronics}, pp. \bibinfo{pages}{37--40}.
\bibitem[{Nakai et~al.(2013)Nakai, Nakagawa, Kuroda, Fujii, Nakamura, and
  Yusa}]{Nakai2013}
\bibinfo{author}{K.~Nakai}, \bibinfo{author}{H.~Nakagawa},
  \bibinfo{author}{K.~Kuroda}, \bibinfo{author}{S.~Fujii},
  \bibinfo{author}{Y.~Nakamura}, \bibinfo{author}{S.~Yusa},
\newblock \bibinfo{title}{Near-infrared-responsive liquid marbles stabilized
  with carbon nanotubes},
\newblock \bibinfo{journal}{Chemistry Letters} \bibinfo{volume}{42}
  (\bibinfo{year}{2013}) \bibinfo{pages}{719--721}.
\bibitem[{Sun et~al.(2012)Sun, Chen, Yang, and Peng}]{Sun2012}
\bibinfo{author}{X.~Sun}, \bibinfo{author}{T.~Chen}, \bibinfo{author}{Z.~Yang},
  \bibinfo{author}{H.~Peng},
\newblock \bibinfo{title}{The alignment of carbon nanotubes: An effective route
  to extend their excellent properties to macroscopic scale},
\newblock \bibinfo{journal}{Accounts of Chemical Research} \bibinfo{volume}{46}
  (\bibinfo{year}{2012}) \bibinfo{pages}{539--549}.
\bibitem[{Cellot et~al.(2009)Cellot, Cilia, Cipollone, Rancic, Sucapane,
  Giordani, Gambazzi, Markram, Grandolfo, Scaini et~al.}]{cellot2009carbon}
\bibinfo{author}{G.~Cellot}, \bibinfo{author}{E.~Cilia},
  \bibinfo{author}{S.~Cipollone}, \bibinfo{author}{V.~Rancic},
  \bibinfo{author}{A.~Sucapane}, \bibinfo{author}{S.~Giordani},
  \bibinfo{author}{L.~Gambazzi}, \bibinfo{author}{H.~Markram},
  \bibinfo{author}{M.~Grandolfo}, \bibinfo{author}{D.~Scaini}, et~al.,
\newblock \bibinfo{title}{Carbon nanotubes might improve neuronal performance
  by favouring electrical shortcuts},
\newblock \bibinfo{journal}{Nature nanotechnology} \bibinfo{volume}{4}
  (\bibinfo{year}{2009}) \bibinfo{pages}{126}.
\bibitem[{Poncharal et~al.(1999)Poncharal, Wang, Ugarte, and
  De~Heer}]{poncharal1999electrostatic}
\bibinfo{author}{P.~Poncharal}, \bibinfo{author}{Z.~Wang},
  \bibinfo{author}{D.~Ugarte}, \bibinfo{author}{W.~A. De~Heer},
\newblock \bibinfo{title}{Electrostatic deflections and electromechanical
  resonances of carbon nanotubes},
\newblock \bibinfo{journal}{science} \bibinfo{volume}{283}
  (\bibinfo{year}{1999}) \bibinfo{pages}{1513--1516}.
\bibitem[{Maciel et~al.(2008)Maciel, Anderson, Pimenta, Hartschuh, Qian,
  Terrones, Terrones, Campos-Delgado, Rao, Novotny et~al.}]{maciel2008electron}
\bibinfo{author}{I.~O. Maciel}, \bibinfo{author}{N.~Anderson},
  \bibinfo{author}{M.~A. Pimenta}, \bibinfo{author}{A.~Hartschuh},
  \bibinfo{author}{H.~Qian}, \bibinfo{author}{M.~Terrones},
  \bibinfo{author}{H.~Terrones}, \bibinfo{author}{J.~Campos-Delgado},
  \bibinfo{author}{A.~M. Rao}, \bibinfo{author}{L.~Novotny}, et~al.,
\newblock \bibinfo{title}{Electron and phonon renormalization near charged
  defects in carbon nanotubes},
\newblock \bibinfo{journal}{Nature materials} \bibinfo{volume}{7}
  (\bibinfo{year}{2008}) \bibinfo{pages}{878}.
\bibitem[{Matsunaga et~al.(2011)Matsunaga, Matsuda, and
  Kanemitsu}]{matsunaga2011observation}
\bibinfo{author}{R.~Matsunaga}, \bibinfo{author}{K.~Matsuda},
  \bibinfo{author}{Y.~Kanemitsu},
\newblock \bibinfo{title}{Observation of charged excitons in hole-doped carbon
  nanotubes using photoluminescence and absorption spectroscopy},
\newblock \bibinfo{journal}{Physical review letters} \bibinfo{volume}{106}
  (\bibinfo{year}{2011}) \bibinfo{pages}{037404}.
\bibitem[{Gale(2019)}]{Gale2019}
\bibinfo{author}{E.~Gale},
\newblock \bibinfo{title}{Neuromorphic computation with spiking memristors:
  habituation, experimental instantiation of logic gates and a novel
  sequence-sensitive perceptron model},
\newblock \bibinfo{journal}{Faraday Discussions} \bibinfo{volume}{213}
  (\bibinfo{year}{2019}) \bibinfo{pages}{521--551}.
\bibitem[{Diederich et~al.(2018)Diederich, Bartsch, Kohlstedt, and
  Ziegler}]{Diedrich2018}
\bibinfo{author}{N.~Diederich}, \bibinfo{author}{T.~Bartsch},
  \bibinfo{author}{H.~Kohlstedt}, \bibinfo{author}{M.~Ziegler},
\newblock \bibinfo{title}{A memristive plasticity model of voltage-based stdp
  suitable for recurrent bidirectional neural networks in the hippocampus},
\newblock \bibinfo{journal}{Scientific Reports} \bibinfo{volume}{8}
  (\bibinfo{year}{2018}).
\bibitem[{Adamatzky(2002)}]{adamatzky2002cbc}
\bibinfo{editor}{A.~Adamatzky} (Ed.), \bibinfo{title}{{Collision-Based
  Computing}}, \bibinfo{publisher}{Springer-Verlag}, \bibinfo{address}{London},
  \bibinfo{year}{2002}.

\end{thebibliography}

\end{document}